\documentclass[twocolumn, superscriptaddress,amsmath,amssymb,aps,prl,floatfix]{revtex4-1}

\usepackage{graphicx}
\usepackage{physics}
\usepackage{comment}
\usepackage{color}
\usepackage{bm}
\usepackage{svg}
\usepackage[colorlinks,urlcolor=blue,citecolor=blue,linkcolor=blue]{hyperref}
\usepackage{cleveref}

\newcommand{\up}{\uparrow}
\newcommand{\down}{\downarrow}

\renewcommand{\k}{{\bf k}}
\newcommand{\p}{{\bf p}}

\newcommand{\q}{{\bf q}}

\newcommand{\0}{{\bf 0}}
\newcommand{\ef}{\varepsilon_F}
\newcommand{\kf}{k_F}

\newcommand{\elp}{E_{LP}}
\newcommand{\dg}{{\dagger}}
\newcommand{\nn}{\nonumber}
\newcommand{\beq}{\begin{equation}}
\newcommand{\eeq}{\end{equation}}

\newcommand{\dn}{\downarrow}
\newcommand{\sch}{Schr{\"o}dinger }

\usepackage{amsmath, url, graphicx,float, braket, simpler-wick, hyperref, amsfonts,amsthm,bm, bbm}
\usepackage{graphicx}
\usepackage{dcolumn}
\usepackage{bm}
\begin{document}

\title{Resonantly enhanced polariton-mediated superconductivity in a doped 
transition metal dichalcogenide monolayer}

\author{Kenneth Choo}
\affiliation{School of Physics and Astronomy, Monash University, Victoria 3800, Australia}
\author{Olivier Bleu}
\affiliation{School of Physics and Astronomy, Monash University, Victoria 3800, Australia}
\affiliation{Institut f{\"u}r Theoretische Physik, Universit{\"a}t Heidelberg, 69120 Heidelberg, Germany}
\author{Meera M. Parish}
\affiliation{School of Physics and Astronomy, Monash University, Victoria 3800, Australia}
\author{Jesper Levinsen}
\affiliation{School of Physics and Astronomy, Monash University, Victoria 3800, Australia}

\date{\today}
\begin{abstract}
We present a proposal for achieving light-induced superconductivity using exciton polaritons---hybrid light-matter particles of excitons (bound electron-hole pairs) and microcavity photons. In contrast to previous theories of polariton-mediated superconductivity, which typically require multiple semiconductor layers, we show that superconductivity can be induced within a single semiconductor monolayer with inverted conduction bands, such as in the tungsten-based transition metal dichalcogenides. The key ingredient is that we can resonantly excite exciton polaritons into bands that are different from those occupied by the doped electrons, thus avoiding any Pauli blocking effects. Crucially, we can exploit the trion fine structure (i.e., multiple exciton-electron bound states) and tune the electron-polariton interactions via Feshbach resonances. Our theory of polariton-mediated superconductivity includes the energy dependence of the polariton-mediated interactions between electrons, as well as the polariton-induced changes to the electron quasiparticles. We find that superconductivity at elevated temperatures is within reach of current experiments.   
\end{abstract}

\maketitle

The application of light to manipulate or induce phases of matter presents an exciting frontier in current research~\cite{BlochReview2022}. 
Already, experiments have made remarkable progress in this direction, with the observations of an optically enhanced magnetic response~\cite{Disa2020}, photo-induced ferroelectric order~\cite{Nova2019,Li2019}, and the transient enhancement of superconductivity at elevated temperatures~\cite{Fausti2011,isoyama2021, sobolev2021}. These recent advances naturally suggest the tantalising possibility of achieving light-controlled superconductivity in the steady state.

A promising direction for this pursuit is that of exciton-polariton-mediated superconductivity in semiconductor microcavities~\cite{laussy2010,laussy2012,cotlet2016,cherotchenko2016,plyashechnik2023}. Here, rather than phonons providing the ``glue'' for electron pairing, as in conventional superconductors~\cite{SchriefferBook}, the necessary attraction is mediated by excitons (bound electron-hole pairs) strongly coupled to cavity photons. 
Such exciton polaritons are hybrid part-light, part-matter quasiparticles, making them ideal for hosting collective quantum phenomena~\cite{carusotto2013review}, including superfluidity at room temperature~\cite{Lerario2017}.  
Most recently, transition metal dichalcogenides (TMDs) have emerged as a promising two-dimensional (2D) material for investigating exciton-polariton physics due to their strong light-matter coupling and strong electronic correlations~\cite{Low2017_review}. In particular, TMDs host robust excitons and trions (exciton-electron bound states) that can survive at room temperature~\cite{wang2018}. This has led to a resurgence in interest in both polariton- and exciton-induced superconductivity~\cite{cotlet2016,julku2022,plyashechnik2023,vonmilczewski2024,zerba2024,Bighin2025}. However, an experimental observation of this exotic superconductor still remains elusive.

\begin{figure}[htb]
\includegraphics[width=1\columnwidth]{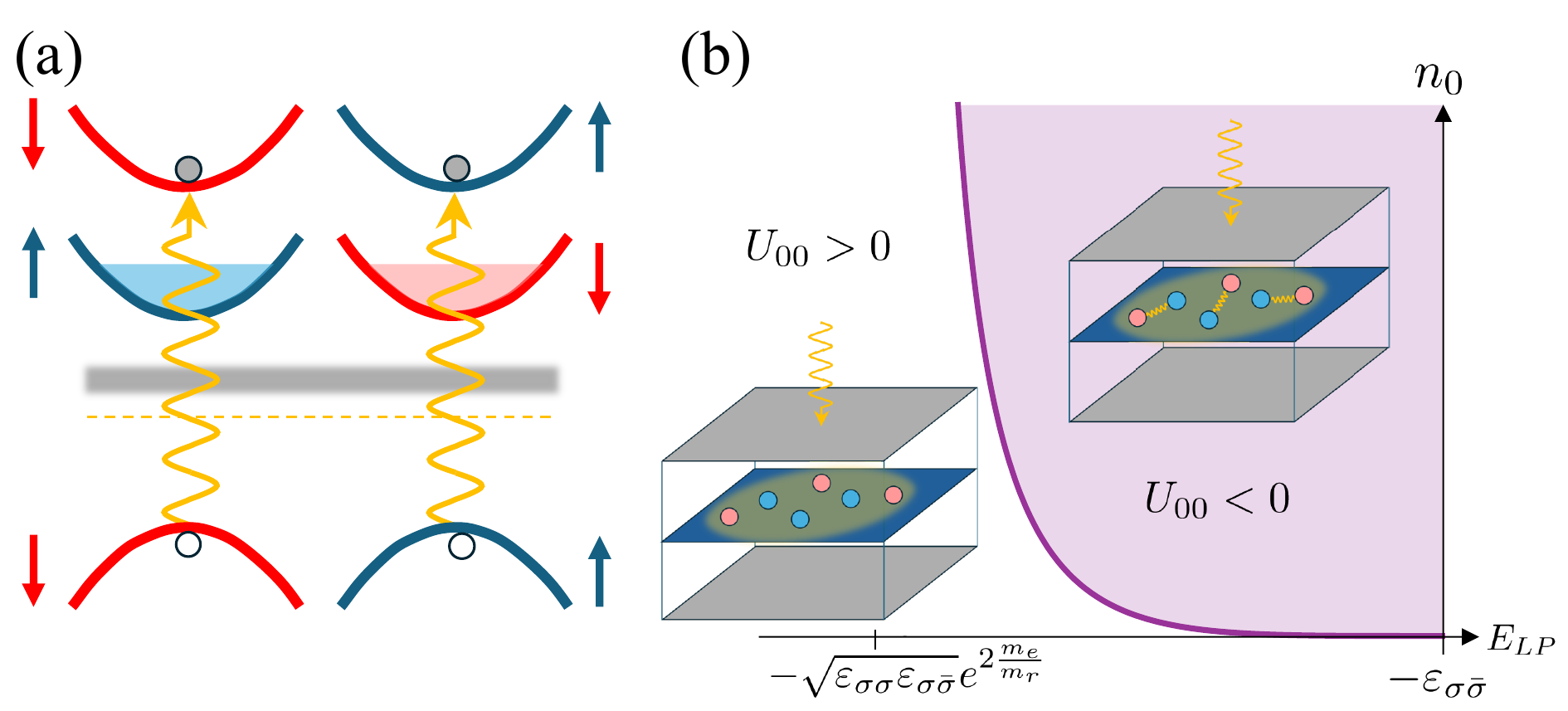}
\caption{(a) The inverted bandstructure of a tungsten-based TMD monolayer allows exciton polaritons to be created in the upper conduction bands, without any Pauli blocking of the doped electrons in the lowest conduction bands. The lower-polariton energy (dashed yellow line) is controlled via the coupling to the cavity photon (wiggly lines) such that it can be tuned below dark states~\cite{Shan2022} (gray region), including trions. (b) The polaritons induce an attraction between $\up$ and $\dn$ electrons which can overcome the repulsive Coulomb interactions and cause the effective electron-electron interaction $U_{\0\0}$ to be net attractive (purple region) when the lower-polariton energy $E_{LP}$ approaches the lowest trion energy $-\varepsilon_{\sigma\bar{\sigma}}$. Far below the trion resonance, there is no attraction for any polariton density $n_0$ once $E_{LP} <-\sqrt{ \varepsilon_{\sigma\sigma}\varepsilon_{\sigma\bar{\sigma}}} \, e^{2(m_e+m_X)/m_X}$ (see text).
The insets illustrate the microcavity setup where the Bose-Fermi mixture of $\up$ and $\dn$ electrons (blue and red circles) and resonantly excited polaritons is realized in a single TMD layer. 
}
\label{fig:illustration}
\end{figure}

A major challenge is the complicated nature of the various proposed geometries, which typically involve multilayer setups where the electrons are separated from the mediating exciton polaritons in another layer~\cite{laussy2010,laussy2012,cotlet2016,cherotchenko2016,plyashechnik2023}. This, in turn, results in substantial screening effects, thus necessitating strong long-range polariton-electron interactions. While such long-range interactions can in principle be achieved by making the excitons dipolar (e.g., using semiconductor quantum wells~\cite{Cristofolini2012, togan2018} or TMD bilayers~\cite{datta2022,Louca2023,sun2024}), this again poses new challenges, as dipolar excitons naturally have a reduced light-matter coupling.

Here we instead propose to realize polariton-induced superconductivity within a single monolayer embedded in a microcavity. 
Specifically, we take advantage of the unique properties 
of TMDs such as WS$_2$ and WSe$_2$, which feature spin-split conduction bands that are inverted [Fig.~\ref{fig:illustration}(a)] due to the strong spin-orbit coupling, as well as a remarkably strong light-matter coupling~\cite{polimeno2025}. 
This has two key advantages: First, the band inversion enables the simultaneous pumping of exciton polaritons in the upper conduction bands and the doping of electrons in the lower conduction bands, without the need to separate the mediating particles and electrons into different layers. Second, the scattering of a polariton and an electron from either valley is near resonant with a trion bound state, and consequently the polariton-induced attraction between electrons can be strongly enhanced [Fig.~\ref{fig:illustration}(b)]. 
This scenario mimics the situation in dilute atomic Fermi gases where the interactions can be resonantly enhanced due to the existence of a bound state~\cite{Bloch2008}, resulting in the highest critical temperature $T_c$ (relative to the Fermi energy) of any system---indeed a similar resonance hypothesis has recently been proposed to explain the elevated critical temperatures in the cuprates~\cite{homeier2025}. Crucially, our proposal can be realized with resonantly driven polaritons since it does not rely on superfluidity of the polariton system nor on the presence of soft modes, and it only requires the ability to embed a gated TMD monolayer in a microcavity, which has already been achieved. We find that the optimal $T_c$ occurs for a range of electron densities and polariton energies that are well within reach of current experiments.

\paragraph{Model.---} We consider the scenario of resonantly excited exciton polaritons interacting with electrons in a band-inverted TMD monolayer such as WX$_2$, which is described by the following Hamiltonian (we set $\hbar$ and area to 1):
\begin{align} \nn
    & \hat{H} =  \sum_{\k, \sigma} \epsilon_{\k }
    \hat{e}^{\dg}_{\k \sigma} \hat{e}_{\k \sigma} + \sum_{\k,\sigma} (
    E_{\k} - \elp)\hat{b}^{\dg}_{\k\sigma} \hat{b}_{\k\sigma} 
    \\
    & + \frac{1}{2} \sum_{\substack{\q, \k, \k' \\\sigma,\sigma'}} \! V_{\q} \, \hat{e}^{\dg}_{\k+\q \sigma} \hat{e}^{\dg}_{\k'-\q \sigma'}\hat{e}_{\k' \sigma'}\hat{e}_{\k \sigma} + \frac{X^2 n_0}{2} \!\!  \sum_{\k, \sigma,\sigma'} g_{\sigma\sigma'} \hat{e}^{\dg}_{\k \sigma} \hat{e}_{\k \sigma} \nn \\
    & + \frac{X\sqrt{n_0}}{\sqrt{2}} \sum_{\substack{\q \neq 0, \\ \k, \sigma,\sigma'}} g_{\sigma\sigma'} \hat{e}^{\dg}_{\k, \sigma} \hat{e}_{\k+\q \sigma} 
 ( \hat{b}^{\dg}_{\q\sigma'} + \hat{b}_{-\q\sigma'} ) 
    \nn \\
    & + \sum_{\substack{\k' \neq \{0, \q\} \\ \k, \q, \sigma, \sigma'}} g_{\sigma\sigma'} \hat{e}^{\dg}_{\k \sigma} \hat{e}_{\k + \q \sigma} \hat{b}^{\dg}_{\k'\sigma'} \hat{b}_{\k'-\q\sigma'} \, . 
    \label{eq:ham}
\end{align}

Here $\hat{e}^{\dg}_{\k \sigma}$ creates electrons with spin $\sigma = \{ \up,\down\}$, momentum $\k$,  and corresponding kinetic energy $\epsilon_{\k} = |\k|^2/2 m_e$, with $m_e$ the electron mass. 
Likewise, $\hat{b}^{\dg}_{\k\sigma}$ creates exciton polaritons with momentum $\k$ and spin $\sigma$ set by the spins of the electron-hole pair [Fig.~\ref{fig:illustration}(a)]. The optical pump is set to the lower polariton energy $\elp$ and it resonantly excites a steady-state condensate density $n_0$ of linearly polarized polaritons at $\k=0$ (normal incidence), such that $\expval*{\hat{b}_{\0\up}} = \expval*{\hat{b}_{\0\down}} = \sqrt{n_0/2}$ (see Supplemental Material~\cite{supmat}).
The energy and associated exciton fraction $X^2$ of the $\k=0$ lower polariton can be controlled via the light-matter coupling strength $\Omega$ and the photon detuning $\delta$ relative to the exciton, with
$\elp  = \frac{1}{2}(\delta - \sqrt{\delta^2 + 4 \Omega^2})$ and $X^2  = \frac{1}{2} \bigg(1+\frac{\delta}{\sqrt{\delta^2 + 4 \Omega^2}}\bigg)$.

Since the typical exciton momenta affected by the light-matter coupling are much smaller than those involved in electron-polariton interactions~\cite{kumar2023}, we approximate the polaritons at finite momentum to be excitons, i.e., $E_\k \simeq |\k|^2/2 m_X$ with exciton mass $m_X$.

The repulsion between the electrons is modelled by an effective screened Coulomb interaction:
\begin{equation} \label{eq:Coulomb}
    V_{\q} = \frac{2 \pi}{m_e (|\q| a_0 + S(\q))},
\end{equation}
where $a_0$ is an effective Bohr radius that characterizes the strength of the Coulomb repulsion~\cite{supmat}.  
The dimensionless function $S(\q) = -2\pi\chi(\q)/m_e$, where $\chi(\q)$ is the static 2D Lindhard function at finite temperature~\cite{vignale2005}. 
Note that we do not consider the Rytova-Keldysh potential~\cite{Rytova1967,Keldysh1979,Cudazzo2011} since the distance between electrons is typically larger than the screening length~\cite{Berkelbach2013} (i.e., $\q$ is sufficiently small) and thus the Coulomb interaction \eqref{eq:Coulomb} is reasonable. 

The electron-exciton interactions are spin-dependent and sufficiently short-ranged~\footnote{The range of  electron-exciton interactions is set by the size of the exciton, which is smaller than all other relevant length scales due to the large exciton binding energy.} that it is accurate to model them with coupling constants $g_{\sigma\sigma'}$ in our Hamiltonian \eqref{eq:ham}, where $g_{\up\up} = g_{\down\down}$ and $g_{\up\down} = g_{\up\down}$. We account for all the interaction terms involving the exciton-polariton condensate, including the usual Fr\"ohlich-style term which resembles the electron-phonon interaction~\cite{Fröhlich1952} that underpins phonon-mediated superconductivity. Crucially, we also have terms beyond the Fr\"ohlich model [last line of Eq.~\eqref{eq:ham}] which produce repeated scattering between electrons and polaritons excited from the condensate. This allows us to capture the trion bound states that are known to exist in WX$_2$ monolayers~\cite{Plechinger2016}, corresponding to pairs of triplet ($\sigma = \sigma'$) and singlet ($\sigma \neq \sigma'$) trions, with binding energies $\varepsilon_{\sigma \sigma'}$. 

\begin{figure}[htb]
\includegraphics[width=1.02\columnwidth]{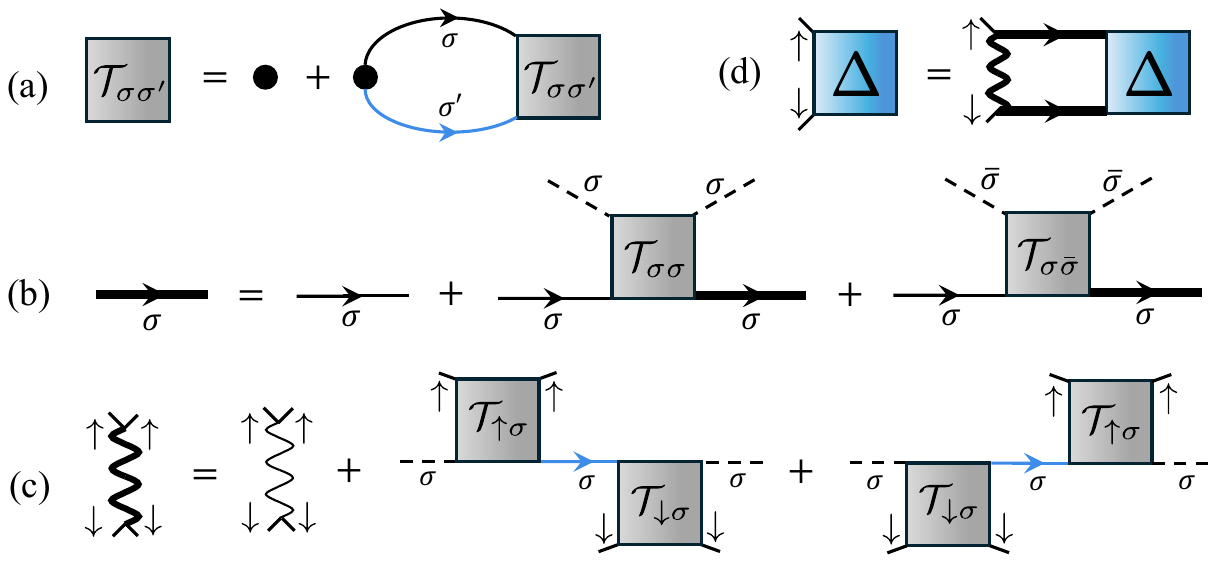}
\caption{Feynman diagrams for polariton-mediated superconductivity. Counter-clockwise from top left: (a) The polariton-electron $T$ matrix $\mathcal{T}_{\sigma \sigma'}$ in Eq.~\eqref{eq:tmatrix} is obtained from all two-body scattering processes between a polariton (blue propagator) and an electron (black propagator)  with interaction $g_{\sigma\sigma'}$ (black dot). (b)  The full electron Green's function (bold line) in Eq.~\eqref{eq:green} obeys the Dyson equation. The self energy involves repeated scattering with polaritons of the same and opposite spin ($\sigma$ and $\bar\sigma$), where the dashed line correspond to $\sqrt{n_0/2}$.
(c)  The total interaction between doped electrons in Eq.~\eqref{eq:potential} (bold wiggly line) involves the repulsive Coulomb interaction (wiggly line), and a polariton-induced attraction at energy $2\mu$, which is summed over the polariton spin $\sigma$. 
(d) The Thouless criterion \eqref{eq:gapeq} that determines the critical temperature $T_c$ for superconductivity, with pairing order parameter $\Delta$.}
\label{fig:feynman}
\end{figure}

All the properties of polariton-electron interactions are encoded in the two-body scattering $T$ matrix obtained from the infinite Born series depicted in Fig.~\ref{fig:feynman}(a):
\begin{align} \label{eq:series}
   \mathcal{T}_{\sigma \sigma'}(\k, \omega) & = g_{\sigma\sigma'}+ g_{\sigma\sigma'}  \Pi(\k,\omega)    \mathcal{T}_{\sigma \sigma'}(\k, \omega) \, ,
 \end{align}
where $\Pi(\k,\omega)$ is the (spin-independent) one-loop polarization bubble 
\begin{equation} \notag
    \Pi(\k,\omega) = i\int \frac{d\omega'}{2\pi} \sum_{\k'} G_e^{(0)}(\k+\k',\omega+\omega') G_P(-\k',-\omega') \, ,
\end{equation}
at energy $\omega$, with the finite-momentum polariton Green's function $G_P(\k,\omega) = (\omega - E_\k +\elp)^{-1}$ and the vacuum electron propagator $G_e^{(0)}(\k,\omega) = (\omega - \epsilon_\k)^{-1}$.

In TMD monolayers, where the collision energy (here dominated by $\elp$) is much smaller than the exciton binding energy, 
we have the low-energy expression~\cite{kumar2023}
 \begin{align}\label{eq:tmatrix}
      \mathcal{T}_{\sigma \sigma'}(\k, \omega) & = \frac{2 \pi}{m_r}\frac{1}{\ln (-\frac{\varepsilon_{\sigma \sigma'}}{\omega+\elp-|\k|^2/2M})} \, ,
\end{align}
with total exciton-electron mass $M = m_e + m_{X}$ and reduced mass 
$m_r = (1/m_e + 1/m_{X})^{-1}$. Here we can see that $\mathcal{T}_{\sigma \sigma'}$ diverges when $\elp$ is close to the trion energy $-\varepsilon_{\sigma \sigma'}$, signifying resonantly enhanced polariton-electron interactions. Indeed, such trion Feshbach resonances are the origin of the observed exciton polaron polaritons in doped TMD monolayers~\cite{Sidler2017}. 

\paragraph{Effect of a polariton condensate on electrons.---} 
We now turn to the many-body system and consider how an electron gas in the monolayer is affected by a (resonantly pumped) polariton condensate. This can be described using the interacting electron Green's function, which at finite temperature takes the form
\begin{equation} \label{eq:green}
    G_{e\sigma}(\k,i\omega_n) = \frac{1}{i\omega_n-\epsilon_{\k} + \mu - \Sigma_\sigma(\k,i\omega_n)}  \, ,
\end{equation}
where $\omega_n = (2n+1) \pi/\beta$ are the fermionic Matsubara frequencies with $\beta = 1/k_BT$, and $\mu$ is the electron chemical potential. The interactions with the polariton condensate are encoded in the self energy $\Sigma_\sigma(\k,i\omega_n)$, which becomes spin independent when the polariton condensate is linearly polarized. As depicted in Fig.~\ref{fig:feynman}(b), 
the dominant interaction process involves repeated polariton-electron scattering, which yields (dropping the spin index)
\begin{equation}
 \Sigma(\k,i\omega_n) = \frac{1}{2} 
    X^2 n_0 (\mathcal{T}_{\up \up}(\k, i\omega_n) + \mathcal{T}_{\up \down}(\k, i\omega_n)) \, .
\end{equation}
Since the interactions only involve the matter component, this scales with the exciton density $X^2n_0$ of the polariton condensate. The leading order effect of the polaritons on the electrons is then the mean-field energy shift $ \Sigma(\0,0)$, thus resulting in an electron Green's function with an effective chemical potential $\tilde{\mu} \equiv \mu - \Sigma(\0,0)$. This in turn is related to the total density of electrons $n_e$ via
\begin{equation}
    n_{e} = \frac{1}{\beta}\sum_{\k,n,\sigma} \!G_{e\sigma}(\k,i\omega_n) \simeq \sum_{\k} \frac{2}{\text{ exp}[\beta(\epsilon_{\k}-\tilde{\mu})] + 1} \, .
\label{eq:numbereq}
\end{equation}

The polariton condensate also mediates an attraction between the $\up$ and $\down$ electrons via finite-momentum polaritons that have been scattered out of the condensate. This process scales as $\mathcal{T}^2$ at leading order in the polariton-electron interactions, as shown in Fig.~\ref{fig:feynman}(c), and yields the total interaction for incoming and outgoing electrons at $(\k\up,-\k\down)$ and $(\k'\up,-\k'\down)$, respectively: 
\begin{equation}
    U_{\k \k'} = V_{\k-\k'} + \frac{X^2 n_0 (\mathcal{T}_{\k \up \up }\mathcal{T}_{\k'\down \up} + \mathcal{T}_{\k'\up \up}\mathcal{T}_{\k\down \up})}{2 \mu - \epsilon_{\k} - \epsilon_{\k'} -E_{\k-\k'} + \elp}  \, ,
\label{eq:potential}
\end{equation}
where $\mathcal{T}_{\k \sigma \sigma'} \equiv  \mathcal{T}_{\sigma \sigma'}(\k, -\epsilon_\k)$. The effective potential $U_{\k\k'}$ is also consistent with the weak-coupling limit of a variational wave function approach involving a pair of electrons scattering with an excitation of the polariton condensate~\cite{supmat}. Note that we explicitly include the Coulomb interaction in Eq.~\eqref{eq:potential} but not in Eq.~\eqref{eq:green}, since it can strongly affect pairing correlations~\cite{supmat} while it primarily produces shifts in the single-particle energies that can be absorbed into the chemical potential.  

The polariton-induced interaction in the second term of Eq.~\eqref{eq:potential} is always attractive since $E_{LP} +2\mu < 0$ and $\mathcal{T}_{\k \up \up }\mathcal{T}_{\k'\down \up} >0$ for the regime we consider. Here we have not retained the full energy dependence of the induced interaction, since we are interested in the leading order behavior. We have instead fixed the energy to be $2\mu$, which corresponds to the bottom of the two-electron scattering continuum and is thus the appropriate energy for determining the critical temperature at which pairs of electrons unbind. The propagator for the finite-momentum polariton excitation is then $G_P(\k-\k', 2\mu - \epsilon_{\k} - \epsilon_{\k'})$ due to energy and momentum conservation, yielding the denominator in Eq.~\eqref{eq:potential}.
This represents a key difference from previous works \cite{laussy2010, laussy2012, cherotchenko2016}, which consider potentials of a similar form to that in Fig.~\ref{fig:feynman}(c), but instead use the kinetic energies of the electrons, which effectively corresponds to perturbing around the unbound scattering states rather than starting from the fundamentally non-perturbative bound pair state. 

Note that we ignore the higher-order effect of a finite electron density on the polariton-electron $T$ matrix, which is reasonable when $|E_{LP}|$ is larger than the Fermi energy $\varepsilon_{F} = \pi n_e/m_e$. 
This condition can easily be fulfilled in TMD monolayers, where electronic densities correspond to Fermi energies ranging from a few meV to around 10 meV~\cite{Sidler2017,koksal2021,ni2025}, while $|E_{LP}|$ in monolayer $\text{WS}_2$ and $\text{WSe}_2$ can be as large as 30-60 meV \cite{krol2019, polimeno2025}. We also neglect the effect of the electrons on the polaritons themselves, since this corresponds to a higher order correction in Eq.~\eqref{eq:potential}, where the leading order effect is a shifted detuning $\delta$. 

Figure~\ref{fig:illustration}(b) illustrates the behavior of the $\up\down$ electron-electron interaction in the limit of vanishing momentum and zero temperature. We see that the induced attraction dominates over the Coulomb repulsion even far below the trion resonance. In particular, for TMD monolayers where $m_e \approx m_X/2$, we have net attraction for $\elp \gtrsim - 20 \sqrt{ \varepsilon_{\sigma\sigma}\varepsilon_{\sigma\bar{\sigma}}}$, which covers the range of energies accessible in experiment. 
As $E_{LP}$ approaches the lowest trion energy $-\varepsilon_{\sigma\bar{\sigma}}$, we see that the polariton density $n_0$ required to achieve electron-electron attraction becomes vanishingly small due to the resonant polariton-electron scattering. To describe the behavior near the trion resonance, one eventually requires a strong-coupling theory that includes the energy dependence of the interactions as well as higher-order electron-polariton scattering processes beyond $\mathcal{T}^2$. 

Finally, to determine the critical temperature $T_c$ below which superconductivity occurs, we consider the condition for a zero-momentum electron pair to bind at energy $2\mu$---the so-called Thouless criterion~\cite{thouless1960}---as depicted in Fig.~\ref{fig:feynman}(d). Performing the sum over Matsubara frequencies in $\sum_{\k',n} U_{\k\k'} G_{e\up}(\k',i\omega_n) G_{e\down}(-\k',-i\omega_n)$ then gives
\begin{equation}
    \Delta_{\k} = -\sum_{\k'} U_{\k \k'} \frac{\Delta_{\k'}}{2 (\epsilon_{\k'}-\tilde{\mu})} \text{tanh }\bigg(\frac{\epsilon_{\k'}-\tilde{\mu}}{2 k_B T_c}\bigg) \, ,
\label{eq:gapeq}
\end{equation}
where $\Delta_{\k}$ corresponds to the pairing order parameter. In the following, we specialize to the case of $s$-wave pairing. 

\begin{figure}[tbh]
\includegraphics[width=1\columnwidth]{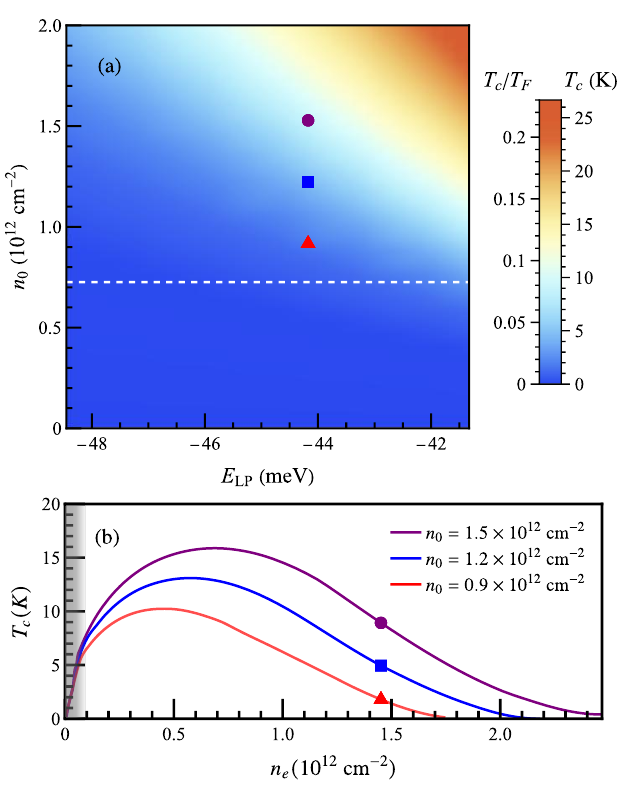}
\vspace{-15pt}
\caption{(a) Critical temperature as a function of the energy and density of the lower polariton, at fixed electron doping $n_e \simeq 1.45 \times 10^{12}\,\text{cm}^{-2}$  (Fermi energy $\varepsilon_F=10\,\text{meV}$) and light-matter coupling strength $\Omega=20 \text{ meV}$. Above the white dashed line, the condition $n_{e}/2 \leq n_0$ is satisfied. (b) Critical temperature as a function of electron doping, for a constant lower polariton energy $E_{LP} = -44.2 \text{ meV}$, and at three fixed polariton densities. The purple circle, blue square and red triangles are the same points between both subfigures. The gray region represents electronic densities for which we expect that Wigner crystallisation \cite{zarenia2017} can compete with superconductivity. In both panels, we have taken the singlet and triplet trion binding energies $\varepsilon_{\sigma\sigma} = 25 \,\text{meV}$ and $\varepsilon_{\sigma\bar\sigma} = 32 \,\text{meV}$ \cite{lyons2019,liu2021}. We also have $a_0=8.5\times10^{-10} \,\text{m}$, while $m_e=0.35m_0$ in terms of the bare electron mass $m_0$ \cite{Goryca2019}. 
}
\label{fig:tc}
\end{figure}

\paragraph{Polariton-mediated superconductivity.---} Figure~\ref{fig:tc}(a) shows our calculated superconducting critical temperature $T_c$ as a function of polariton energy and density, where for concreteness we use parameters for a WSe$_2$ monolayer (see caption). We keep the electron density fixed with $k_B T_F \equiv\ef=10\,\text{meV}$, which is sufficiently below the spin-splitting of the conduction bands~\cite{liu2015, Kormanyos2015} that doping only minimally affects polariton formation. As expected, we observe that $T_c$ increases both when the polariton density is increased and when the polariton energy approaches the singlet trion energy $-\varepsilon_{\up\down}$, i.e., in the vicinity of the trion scattering resonance. Crucially, our calculations indicate that critical temperatures on the order of 10\,K may be obtained for electron doping and polariton densities on the order of $10^{12}\,\text{cm}^{-2}$, densities which are already achieved in TMDs.

In Fig.~\ref{fig:tc}(b), we instead examine the effect of varying the electron doping at fixed polariton density. In the limit $n_e \to 0$, the effective attraction is suppressed and thus $T_c$ vanishes, similarly to the mean-field behavior of a 2D Fermi gas, where $T_c \sim \sqrt{\ef}$~\cite{levinsen2015}. Furthermore, our theory likely overestimates $T_c$ in the low-density regime, since it does not include the possibility of Wigner crystallization which is expected to dominate in this limit~\cite{zarenia2017}; indeed, signatures of electronic crystalline order have recently been observed in the related TMD MoSe$_2$~\cite{smolenski2021}. 
In the opposite regime of large electron density, where the repulsive Coulomb interaction is screened, we expect $T_c$ to again be suppressed once $n_0 < n_e/2$ because then there are too few polaritons to mediate a sufficient attraction between $\up$ and $\down$ electrons, i.e., fewer than one polariton per electron pair. This suppression is clearly visible in Fig.~\ref{fig:tc} and is connected to the polariton-induced energy shift of the electrons within our theory~\cite{supmat}.
Thus, we find that the maximal $T_c$ for superconductivity is achieved in the intermediate doping regime, where the optimal electron density is approximately $n_e\approx n_0/2$.

It is instructive to compare our results with the fermionic superfluidity (the analogue of superconductivity) observed in 2D ultracold atomic gases~\cite{levinsen2015}. Here, similar relative critical temperatures $T_c\gtrsim0.1T_F$ (corresponding to tens of nK in that system) are routinely achieved~\cite{feld2011, sommer2012, makhalov2014, ries2015, sobirey2021}. The key ingredient is precisely the ability to strongly enhance the direct interactions between different atomic species close to Feshbach resonances~\cite{Chin2010}. In our scenario, strong attraction between electrons is instead accomplished  by ensuring that the scattering with the mediating polaritons is near resonant.

\textit{Conclusions and outlook}\textbf{---} 
To conclude, we have proposed a simple setup for realizing polariton-mediated superconductivity, which relies on the separation of electrons and polaritons between different bands rather than between different semiconductor layers. This allows one to achieve tunable polariton-electron scattering via the multiple trion bound states. 
Using a theory that includes the polariton-induced energy shift of each electron as well as the Coulomb repulsion and polariton-mediated attraction between electrons, we have obtained superconducting critical temperatures of order 10\,K for parameters in current TMD experiments. An interesting future extension would be to vary the polarization/spin of the polaritons and thus achieve a light-induced effective Zeeman field on the electrons.

Our current work has focussed on rather negative cavity photon detunings [i.e., $X^2=0.14$ to 0.18 from left to right in Fig.~\ref{fig:tc}(a)], since this ensures that we avoid the region close to resonance where our perturbative theory becomes less reliable. However, from a practical point of view, in order to achieve stable light-induced superconductivity, one just requires the lower polariton energy to be below the trion bound states, such that there are no nearby lower-lying (dark) states for the electrons to decay into. Thus, we would encourage experiments to explore the region close to the trion resonance since this is likely to yield the largest critical temperatures.

\acknowledgments 
\textit{Acknowledgements}\textbf{---}We acknowledge useful discussions with Dmitry Efimkin, Richard Schmidt, Tilman Enss, Atac Imamoglu, St\'ephane K\'ena-Cohen and Alexey Sokolik. We acknowledge support from the Australian Research Council (ARC) Centre of Excellence in Future Low-Energy Electronics Technologies (CE170100039).  JL and MMP are also supported through the ARC Discovery Project DP240100569 and ARC Future Fellowship FT200100619, respectively, and jointly through the ARC Discovery Project
DP250103746.
OB also acknowledges support from the Deutsche Forschungsgemeinschaft (DFG) via the Collaborative Research Centre SFB 1225 ISOQUANT (Project-ID No. 273811115). 
KC acknowledges support from an Australian Government Research Training Program (RTP) Scholarship.
\bibliography{superbib}

%%%%%%%%%%%%%%%%%%%%%%%%%%%%%%%%%%%%%%%%%%%%%%%%%%%%
\clearpage
\renewcommand{\theequation}{S\arabic{equation}}
\renewcommand{\thefigure}{S\arabic{figure}}
\onecolumngrid
\newpage
\setcounter{equation}{0}
\setcounter{figure}{0}
\setcounter{page}{1}
%%%%%%%%%%%%%%%%%%%%%%%%%%%%%%%%%%%%%%%%%%%%%%%%%%%%

\begin{center}
\textbf{\large Supplemental Material: Resonantly enhanced polariton-mediated superconductivity in a doped 
transition metal dichalcogenide monolayer}\\
\vspace{4mm}
{K. Choo,$^1$
O. Bleu,$^{1,2}$
M. M. Parish,$^1$
J. Levinsen$^1$}\\
\vspace{2mm}
{\em \small
$^1$School of Physics and Astronomy, Monash University, Victoria 3800, Australia\\
$^2$ Institut f{\"u}r Theoretische Physik, Universit{\"a}t Heidelberg, 69120 Heidelberg, Germany}\end{center}

\section{Model}
We start from the Hamiltonian $\hat{H} = \hat{H}_0 + \hat{V}_{ep}+\hat{V}_{ee}$ with
\begin{subequations}
\begin{align}
    \hat{H}_{0} &= \sum_{\k, \sigma} \epsilon_{\k }\hat{e}^{\dg}_{\k \sigma} \hat{e}_{\k \sigma} + \sum_{\k, \sigma} (E_{\k} - \elp)\hat{b}^{\dg}_{\k \sigma} \hat{b}_{\k \sigma}, \\ \label{eq:Vep}
    \hat{V}_{ep}&=  \sum_{\substack{\q, \k, \k' \\ \sigma \sigma'}} g_{\sigma \sigma'} \hat{e}^{\dg}_{\k \sigma} \hat{e}_{\k + \q \sigma} \hat{b}^{\dg}_{\k' \sigma'} \hat{b}_{\k'-\q \sigma'} ,\\ 
  \hat{V}_{ee}&=\frac{1}{2}\sum_{\substack{\q, \k, \k' \\ \sigma \sigma'}} V_{\q} \hat{e}^{\dg}_{\k+\q \sigma} \hat{e}^{\dg}_{\k'-\q \sigma'}\hat{e}_{\k' \sigma'}\hat{e}_{\k \sigma}.
\end{align}
\end{subequations}
As in the main text, we take $\epsilon_\k=k^2/2m_e$ and $E_\k=k^2/2m_X$ and we work in units where the system area and the reduced Planck constant are both set to 1. To account for the presence of the spin-balanced polariton condensate at $\k=0$, we make the replacement $\hat{b}_{\k\sigma}=X_0\sqrt{\frac{n_0}{2}}\delta_{\k0}+\hat{b}_{\k\neq0\sigma}$, which gives
\begin{align}\label{eq:epinterbeyondF}
    \hat{V}_{ep}&=  \frac{X^2 n_0}{2} \sum_{\k, \sigma, \sigma'} g_{\sigma \sigma'} \hat{e}^{\dg}_{\k \sigma} \hat{e}_{\k \sigma} + \frac{X \sqrt{n_0}}{\sqrt{2}} \sum_{\substack{\q \neq 0, \\ \k, \sigma, \sigma'}} g_{\sigma \sigma'} \hat{e}^{\dg}_{\k \sigma} \hat{e}_{\k+\q \sigma} ( \hat{b}^{\dg}_{\q \sigma'} + \hat{b}_{-\q \sigma'} ) + \sum_{\substack{\k' \neq \{0, \q\}, \\ \k, \q, \sigma \sigma'}} g_{\sigma \sigma'} \hat{e}^{\dg}_{\k \sigma} \hat{e}_{\k + \q \sigma} \hat{b}^{\dg}_{\k' \sigma'} \hat{b}_{\k'-\q \sigma'},
\end{align}
thus yielding the Hamiltonian in the main text. We observe that the last term in \eqref{eq:epinterbeyondF} goes beyond the Fr{\"o}hlich Hamiltonian \cite{Fröhlich1954,Tempere2009}, which is typically used to derive boson-mediated electron-electron interactions. As we will see below, the  presence of this term allows us to recover the low-energy polariton-electron $T$-matrix, and is key to the derivation of our effective electron-electron potential.

\section{Alternative derivation of the polariton-mediated attraction}
\label{supp:derivation}

We now present an alternative wave function based derivation of the effective electron-electron interaction potential, complementing the diagrammatic derivation in the main text.
We note that, for a medium at fixed chemical potential such as a resonantly pumped polariton condensate, there is negligible mediated interaction between same-spin identical electrons due to a cancellation between exchange and Hartree processes~\cite{Mora2010}. We therefore focus on the effective interactions between distinguishable electrons.

To understand how the effective electron-electron interactions are modified by the presence of a polariton condensate, we use the following two-electron ansatz
\begin{equation}
    \ket{\psi} = \sum_{\k} \varphi_{\k} \hat{e}^{\dg}_{\k \up}\hat{e}^{\dg}_{-\k \down} \ket{\Phi_{0}} + \sum_{\k_1, \k_2, \sigma} \varphi^\sigma_{\k_1 \k_2} \hat{e}^{\dg}_{\k_1 \up}\hat{e}^{\dg}_{-\k_2 \down} \hat{b}^{\dg}_{\k_2 - \k_1, \sigma}\ket{\Phi_{0}},
\end{equation}
where $\ket{\Phi_0}$ denotes a state with no electrons and a linearly-polarized polariton condensate at zero momentum. The first term corresponds to the two electrons and the unpertubed condensate, while the second term describes the possibility to excite a boson from each species. Note that a similar ansatz was used in \cite{naidon2018} to describe two impurities in a single-component Bose condensate.

By projecting the \sch equation $\hat{H} \ket{\psi} = E \ket{\psi}$ onto the different basis states, we arrive at coupled equations for the coefficients $\varphi_{\k}$ and $\varphi^{\sigma}_{\k_1 \k_2}$:
\begin{subequations}
\begin{align}
  & \left[ E -2\epsilon_\k  - \frac{X^2 n_0}{2} (g_{\up \up}+ g_{\dn \dn} + g_{\up \down}+g_{\dn \up}) \right]\varphi_{\k} = X\sqrt{\frac{n_0}{2}} \sum_{\sigma}  \left[\chi_{\up \sigma}(\k) + \chi_{\dn\sigma}(\k) \right]
   + \sum_{\p} V_{\k-\p} \varphi_\p ,\label{eq:phi} \\ \nonumber
  &   \left[ E -\epsilon_{\k_1}-\epsilon_{\k_2} -E_{\k_2-\k_1} +\elp- \frac{X^2 n_0}{2} (g_{\up \up}+ g_{\dn \dn} + g_{\up \down}+g_{\dn \up}) \right] \varphi^{\sigma}_{\k_1 \k_2} = X\sqrt{\frac{n_0}{2}} g_{\up \sigma}\varphi_{\k_2}+X\sqrt{\frac{n_0}{2}}g_{\dn \sigma}\varphi_{\k_1}
     \\
     & ~~~~~~~~~~~~~~~~~~ ~~~~~~~~~~~~~~~~~~~~~~ ~+\chi_{\up \sigma}(\k_2) +\chi_{\down \sigma}(\k_1) +\sum_{\q} V_{\q}\varphi_{\k_1-\q,\k_2-\q}^\sigma,
     \label{eq:phi12s} 
\end{align}
\end{subequations}
where we have introduced the functions
\begin{subequations}
\begin{align}
    \chi_{\up \sigma}(\k) &= g_{\up \sigma} \sum_{\p} \varphi_{\p \k}^{\sigma},
 \\ \chi_{\dn \sigma}(\k) &= g_{\dn \sigma} \sum_{\p} \varphi_{\k\p}^{\sigma}.
\end{align}
\end{subequations}
The fact that these functions appear in Eq.~\eqref{eq:phi12s} is due to the third term in Eq.~\eqref{eq:epinterbeyondF}.

One can then manipulate Eq.~\eqref{eq:phi12s} (i.e., divide by $[E-...]$ and sum over $\k_1$ or $\k_2$) to obtain
\begin{subequations}
\label{eq:chieq1}
\begin{align}
    \frac{1}{g_{\up\sigma}}     \chi_{\up \sigma}(\k_2) &= \sum_{\k_1} \frac{ X  \sqrt{\frac{n_0}{2}}(g_{\up \sigma} \varphi_{\k_2} + g_{\dn \sigma} \varphi_{\k_1}) +\chi_{\up \sigma}(\k_2) + \chi_{\dn \sigma}(\k_1)  +\sum_{\q} V_{\q}\varphi_{\k_1-\q,\k_2-\q}^\sigma}{E -\epsilon_{\k_1}-\epsilon_{\k_2} -E_{\k_2-\k_1}+\elp - \frac{X^2 n_0}{2} (g_{\up \up}+ g_{\dn \dn} + g_{\up \down}+g_{\dn \up}) },\\
    \frac{1}{g_{\dn\sigma}}     \chi_{\dn\sigma}(\k_1) &= \sum_{\k_2} \frac{ X  \sqrt{\frac{n_0}{2}}(g_{\up \sigma} \varphi_{\k_2} + g_{\dn \sigma} \varphi_{\k_1}) +\chi_{\up \sigma}(\k_2) + \chi_{\dn \sigma}(\k_1)  +\sum_{\q} V_{\q}\varphi_{\k_1-\q,\k_2-\q}^\sigma }{E -\epsilon_{\k_1}-\epsilon_{\k_2} -E_{\k_2-\k_1}+\elp - \frac{X^2 n_0}{2} (g_{\up \up}+ g_{\dn \dn} + g_{\up \down}+g_{\dn \up}) },
\end{align}
\end{subequations}
We can now relate the bare electron-exciton couplings to the trion binding energies $\varepsilon_{\sigma\sigma'}$ using
\begin{equation}\label{eq:renorm}
    \frac{1}{g_{\sigma\sigma'}}  = -\sum_{\k}^\Lambda \frac{1}{\varepsilon_{ \sigma\sigma'} +\epsilon_{\k} + E_{\k}},
\end{equation}
Equation~\eqref{eq:renorm} implies that $g_{\sigma\sigma'}\rightarrow 0^{-}$ when $\Lambda\rightarrow \infty$. This implies that the terms where $g_{\sigma\sigma'}$ is in front of a constant on the RHS of \eqref{eq:chieq1} will vanish, and that we have the limiting behavior $\sum_{\k} \frac{g_{\sigma\sigma'} \varphi_{\k}}{E-...} \rightarrow 0$, and $\sum_{\k} \frac{g_{\sigma\sigma'}}{E-...} \rightarrow 1$ as $\Lambda\rightarrow \infty$. Furthermore, we note that the terms involving the Coulomb potential on the RHS of Eq.~\eqref{eq:chieq1} are sub-leading compared with the bare Coulomb repulsion, and we will therefore neglect these in the following.

The above expressions then simplify to
\begin{subequations}
\label{eq:chieqb}
\begin{align}
 \mathcal{T}_{\k_2 \up\sigma}^{-1}  \chi_{\up \sigma}(\k_2) &= X  \sqrt{\frac{n_0}{2}} \varphi_{\k_2} +\sum_{\k_1} \frac{ \chi_{\dn \sigma}(\k_1)}
 {E -\epsilon_{\k_1}-\epsilon_{\k_2} -E_{\k_2-\k_1}+\elp  },\\
 \mathcal{T}_{\k_1 \dn\sigma}^{-1}      \chi_{\dn\sigma}(\k_1) &=X  \sqrt{\frac{n_0}{2}}  \varphi_{\k_1} +\sum_{\k_2} \frac{  \chi_{\up \sigma}(\k_2)}
 {E -\epsilon_{\k_1}-\epsilon_{\k_2} -E_{\k_2-\k_1}+\elp  }.
\end{align}
\end{subequations}
Here, we have introduced the notation 
\begin{align}
  \mathcal{T}_{\k \sigma\sigma'}\equiv  T_{\sigma\sigma'}(\k, E - \epsilon_{\k}), \label{eq:tmat}
\end{align}
with the electron-polariton $T$-matrix 
\begin{subequations}
\begin{align}
   T_{\sigma\sigma'}( \k,\omega) &= \left[\frac{1}{g_{\sigma\sigma'}} - \sum_{\q} \frac{1}{\omega-\epsilon_{\q} -E_{\q-\k}+ \elp}\right]^{-1}\\
   & = \left[\frac{1}{g_{\sigma\sigma'}} - \sum_{\q} \frac{1}{\omega-\frac{\q^2}{2m_r} -\frac{\k^2}{2M}+ \elp}\right]^{-1} \label{eq:tmat},
\end{align}
\end{subequations}
where in the second line, we have performed the momentum translation $\q\rightarrow \q +\frac{m_e}{M}\k$, with the total mass $M=m_e+m_X$ and the electron-exciton reduced mass $m_r=m_e m_X/(m_e+m_X)$.
We can recognize that Eq.~\eqref{eq:tmat} corresponds to the vacuum electron-polariton $T$ matrix with center of mass momentum $\k$.
Using \eqref{eq:renorm} it can be calculated analytically and reads
\begin{equation}
    T_{\sigma \sigma'}(\k,\omega)= \frac{2 \pi}{m_r}\frac{1}{\ln \left[\frac{-\varepsilon_{\sigma \sigma'}}{\omega+ \elp- \k^2/(2 M)}\right]}.
\end{equation}
We note that the validity of the above simple analytical expression is confirmed in microscopic calculations accounting for the composite nature of the exciton \cite{kumar2023}.

Finally, we evaluate $\chi_{\sigma \sigma'} $ in \eqref{eq:chieqb} up to second order in $\mathcal{T}_{\sigma\sigma'}$, 
and inject the result into Eq.~\eqref{eq:phi}. This gives
\begin{equation}\label{eq:cooper}
    \left[E - 2\epsilon_\k-\Sigma_{\up}(\k,E)-\Sigma_{\dn}(\k,E)\right]\varphi_{\k} =\sum_{\p} U(\k,\p,E)\varphi_\p,
\end{equation}
where we have introduced
\begin{subequations}
\begin{align}\label{eq:selfenergy}
\Sigma_{\sigma}(\k,E)&=\frac{X^2_0 n_0}{2} \sum_{\sigma'}\mathcal{T}_{\k \sigma \sigma'}, \\ \label{eq:effective_potential}
    U(\k,\p,E) & =\frac{X^2 n_0}{2}  \sum_{\sigma }\frac{\mathcal{T}_{\k \up\sigma}\mathcal{T}_{\p\down\sigma} +\mathcal{T}_{\p \up\sigma}\mathcal{T}_{\k\down\sigma}}{E - \epsilon_{\k} - \epsilon_{\p} -E_{\k-\p} + \elp}+V_{\k-\p} .
\end{align}
\end{subequations}
We can see that Eq.~\eqref{eq:cooper} looks like an effective two-body \sch equation for the $\up\dn$ electron pair. The presence of the polariton condensate has two implications. First, looking on the LHS we can observe that it leads to a renormalization of the electron kinetic energies, which can be interpreted as a polaronic shift caused by the polariton medium. Second and most importantly, it modifies the effective electron-electron interaction potential and can give rise to electron-electron \textit{attraction}.

In practice, we fix the collision energy $E$ of the two electrons to be $2\mu$, which corresponds to the bottom of the two-electron scattering continuum. Equation~\eqref{eq:cooper} thus implies that the leading order effect of the polaritons on the electrons is a mean-field energy shift $ \Sigma(\0,0)$ (which is spin independent for a linearly polarized condensate), thus resulting in an effective chemical potential 
$    \tilde{\mu} \equiv \mu - \Sigma(\0,0),$
which should be used to determine the electron occupation. Note that, in order to avoid going beyond the perturbative limit within which we work, we do not include $\mu$ within the self energy, and likewise this is not included within the scattering $T$ matrices in the polariton-induced attraction. Therefore, using the fact that $\mathcal{T}_{\k\up\up}=\mathcal{T}_{\k\dn\dn}$ and $\mathcal{T}_{\k\up\dn}=\mathcal{T}_{\k\dn\up}$, Eqs.~\eqref{eq:selfenergy} and \eqref{eq:effective_potential} reduce to the self-energy and effective potential given in the main text.

To our knowledge, the above derivation and the resulting induced potential are original. An important difference of the present induced potential with respect to those that can be obtained from the Fr{\"o}hlich Hamiltonian via a Schrieffer-Wolff \cite{Fröhlich1952,laussy2010} or Wegner flow \cite{LenzWegner1996,cotlet2016,Bighin2025} transformation is that it involves the polariton-electron $T$ matrices, and thus, it accounts for the enhancement of the polariton-electron scattering in the vicinity of the trion bound states.

\section{Screened Coulomb repulsion}
\label{supp:coulomb}
To model the repulsion between electrons, we use a temperature-dependent screened Coulomb interaction for $V_{\q}$:
\begin{equation}
    V_{\q} = \frac{2 \pi/m_e}{|\q| a_0 + S(\q)},
\end{equation}
where $S(\q) = -2\pi\chi(\q)/m_e$ is proportional to the finite-temperature Lindhard function:
\begin{equation}
    \chi(\q)=\sum_{\k \sigma} \frac{n_{\k}- n_{\k+\q}}{\epsilon_{\k}-\epsilon_{\k+\q} + i0}.
\label{eq:screent}
\end{equation}
Here, $+i0$ corresponds to an infinitesimal positive imaginary shift~\cite{vignale2005}, and $n_{\k}$ is the Fermi-Dirac distribution
\begin{align}
    n_{\k}=\frac{1}{\exp[\beta(\epsilon_{\k}-\tilde{\mu})] + 1}.
\end{align}
We evaluate $S(\q)$ using the method outlined in Ref.~\cite{vignale2005}.

We furthermore approximate the effective Bohr radius $a_0$ via the 2D hydrogenic relation $a_0 = \sqrt{1/(m_e \varepsilon_{X})}$. This is motivated by the fact that the typical momentum explored within the gap equation is set by the Fermi wave vector $\kf$, which corresponds to length scales much larger than the screening length in 2D materials~\cite{Cudazzo2011}. Hence the Rytova-Keldysh potential~\cite{Rytova1967,Keldysh1979} is only probed at small momenta where the Coulombic approximation is appropriate~\cite{chernikov2014}. The validity of this approximation further improves if the monolayer is encapsulated in hBN.

\section{Effect of Coulomb repulsion and polaronic effects}

\begin{figure}[thb]
\includegraphics[width=0.9\columnwidth]{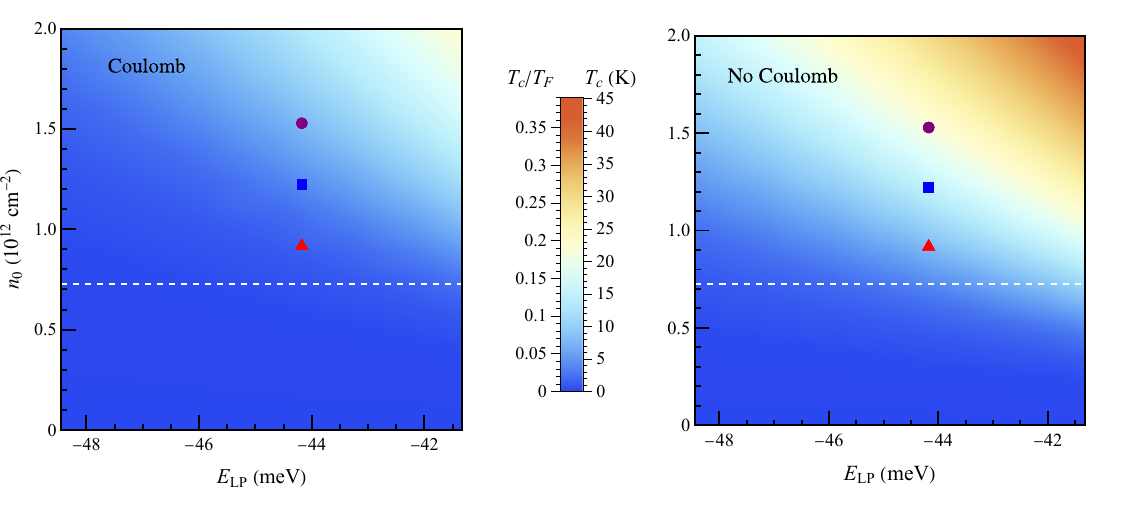}
\caption{Critical temperatures calculated (a) with and (b) without Coulomb repulsion. Panel (a) is the same as in Fig.~3(a) of the main text, with a different color scale.
}
\label{fig:tccompare}
\end{figure}

\begin{figure}[tbh]
\includegraphics[width=0.5\columnwidth]{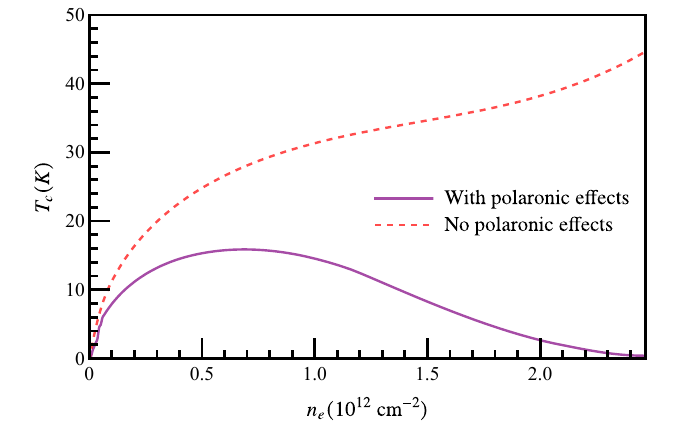}
\vspace{-15pt}
\caption{Critical temperatures as a result of varying the electron doping at fixed polariton density $n_0=1.5 \times 10^{12} \text{ cm}^{-2}$.}
\label{fig:tcpolaron}
\end{figure}

In Figs.~\ref{fig:tccompare} and \ref{fig:tcpolaron} we explore the effects of the electronic Coulomb repulsion and the polaronic dressing of the electrons. First, as can be seen in Fig.~\ref{fig:tccompare}, the result of including direct electron repulsion is a reduction by about a factor of 2 in the critical temperature compared with not including repulsion. However, we stress that the qualitative behavior is the same, with both cases exhibiting enhancements in the critical temperature as we approach the trion resonance and as the polariton density is increased. 

Second, Fig.~\ref{fig:tcpolaron} explores what happens if we do not include the polaronic dressing. Here, we observe a qualitative difference, where in the absence of dressing the critical temperature monotonically increases with increasing doping. 
In that case, we do not capture how there has to be enough polaritons in order to efficiently mediate an attractive interaction between the electrons.

\section{Outline of the numerical approach}
The critical temperatures of polariton-induced superconductivity shown in the figures in the main text and above are extracted by simultaneously solving three equations. These are the number equation,
\begin{equation}
    n_{e} = \sum_{\k} \frac{2}{\text{ exp}[\frac{\epsilon_{\k}-\tilde{\mu}}{k_BT_c}] + 1} \, ,
\label{eq:numbereqSM}
\end{equation}
the polaronic equation
\begin{align}\label{eq:hfmuSM}
    \tilde{\mu} \equiv \mu - \Sigma(\0,0)\,,
\end{align} 
and the Thouless criterion for the order parameter
\begin{equation}
    \Delta_{\k} = -\sum_{\k'} U_{\k \k'} \frac{\Delta_{\k'}}{2 (\epsilon_{\k'}-\tilde{\mu})} \text{tanh }\bigg(\frac{\epsilon_{\k'}-\tilde{\mu}}{2 k_B T_c}\bigg) \, .
\label{eq:gapeqSM}
\end{equation}
Here we have explicitly evaluated the number equation at the critical temperature $T_c$, thus yielding
\begin{equation}
    \tilde{\mu} = k_{B} T_c \ln \left[\exp\bigg(\frac{\varepsilon_{F}}{k_{B} T_c}\bigg) - 1\right]\,,
\end{equation}
with $\varepsilon_F=\frac{\pi n_e}{m_e}$ the Fermi energy.

In practice, we recast the ($s$-wave projected) gap equation as an eigenvalue problem using the method of Gauss-Legendre quadrature~\cite{numericalrecipes}. We manipulate this such that $1/X_0^2n_0$ is the eigenvalue for a given $\mu$, $T_c$, and $n_e$, and we then look for intersections of $1/X_0^2n_0$ with that obtained from the polaronic equation at the same parameter values. We choose the smallest $n_0$ for which all three equations are satisfied.

\end{document}